\documentclass[preprint, prb, aps, superscriptaddress, titlepage, nolongbibliography]{revtex4-2}
\usepackage{amsmath}
\usepackage{bm}
\usepackage{hyperref}
\usepackage{url}
\usepackage{graphicx}
\usepackage{subfigure}
\DeclareMathOperator{\tr}{tr}
\renewcommand{\selectlanguage}[1]{}
\begin{document}
\title{Coexistence of antiferromagnetism and topological superconductivity on the honeycomb lattice Hubbard model}
\date{\today}
\author{Yang Qi}
\thanks{Current address: Department of Physics and State Key Laboratory of Surface Physics, Fudan University, Shanghai 200433, China.}
\affiliation{Department of Physics, Massachusetts Institute of Technology, Cambridge, MA 02138, USA}
\author{Liang Fu}
\affiliation{Department of Physics, Massachusetts Institute of Technology, Cambridge, MA 02138, USA}
\author{Kai Sun}
\affiliation{Department of Physics, University of Michigan, Ann Arbor, MI 48109, USA}
\author{Zhengcheng Gu}
\affiliation{Department of Physics, The Chinese University of Hong Kong, Shatin, New Territories, Hong Kong}

\begin{abstract}
Motivated by the recent numerical simulations for doped
$t$-$J$ model on the honeycomb lattice, we study superconductivity of
singlet and triplet pairing on the honeycomb lattice Hubbard model. We show that a superconducting state with coexisting spin-singlet and spin-triplet pairings is induced by the antiferromagnetic order near half filling. The superconducting state we obtain has a topological phase transition that separates a topologically trivial state and a nontrivial state with Chern number two. Possible experimental realization of such a topological superconductivity is also discussed.
\end{abstract}
\date{September 24, 2017}
\maketitle

\section{Introduction}
\label{sec:introduction}

Antiferromagnetism and superconductivity are two key phenomena that
appear in high temperature superconductors such as cuprates and
iron pnictides~\cite{anderson87, Baskaran1987, Demler2001,
  LNW2006Review, SO5Review, Kamihar2008, AF1, AF2,AF3, AF4, IronAge, JDaiIronPNAS2009}. In these
systems, interaction creates strong magnetic correlations between
electrons and leads to a Mott insulator with antiferromagnetic (AFM)
order for undoped cuprates and a bad metal with spin density wave
(SDW) order for undoped iron pnictides. Upon doping, the magnetic
order disappears and superconductivity (SC) takes place. There have
been many discussions on the roles played by these two different
orders in the phase diagram. On one hand, it has been argued that
magnetic fluctuations play an essential role for the mechanism of high
temperature superconductivity, especially in a class of theory based
on the novel concept of spin-charge separation and RVB
scenario~\cite{anderson87, Baskaran1987, Zou1988, LNW2006Review},
where the metastable spin liquid state(which has a short-range AFM
order and is energetically close to the AFM state) naturally leads to
SC order upon doping.  On the other hand, the concept of
quantum criticality suggests that the AFM order or the SDW order is a
competing order that suppresses SC order~\cite{Demler2001,
  ZhangCO2002, Kivelson2002, SachdevRMP2003, Moon2009,
  She2010, BlackSchaffer2015}. Although the strongly coupling pictures seem to be very
elegant and attractive, so far there is no controlled way to perform
microscopic calculations starting from realistic models, e.g., Hubbard
model with strong repulsive interactions. Therefore, to understand the
interplay between AFM order and SC order is still
an open question and it plays a crucial role for understanding the
underlying physics in these systems.

In this paper, we propose an effective Ginzburg-Landau theory to study the interplay between AFM order and
SC order in the honeycomb lattice Hubbard model, which has
been intensively studied recently. At half filling, antiferromagnetism
in the undoped honeycomb lattice has been studied using quantum Monte
Carlo and other analytical methods~\cite{Sorella1992, Furukawa2001,
  Singh2010}. In these studies an AFM phase is found
above a critical on-site repulsion $U_c$. Upon doping,
SC order has been found in the doped model using various
methods~\cite{Uchoa2007, Kuroki2008, Wu2013, Black-Schaffer2007,
  Pathak2010}, where different pairing symmetries have been found, including
spin-singlet $s$-wave, $d+id$-wave pairing and spin-triplet $p$-wave,
$f$-wave pairing.


In a recent Grassmann tensor product state(GTPS) numerical study of the honeycomb lattice $t$-$J$
model~\cite{Gu2011}, a phase with coexisting AFM and SC
orders has been found at low doping levels. Particularly, the
superconducting state that coexists with AFM order has both
$d+id$ spin-singlet and $p+ip$ spin-triplet pairings. However, the GTPS numerical study could not tell us whether the $d+id$/$p+ip$ SC state is topologically trivial or nontrivial, since the numerical results can not distinguish strong pairing and weak pairing cases. We find that the proposed Ginzburg-Landau theory
can naturally explain such a result based on the trilinear term which naturally couples AFM, $d+id$ spin-singlet pairing and $p+ip$ spin-triplet pairing. Moreover, the proposed trilinear term also suggests a topological
  phase transition that separates a topologically trivial state and a
  nontrivial state with Chern number two.
Although the microscopical origin of such a trilinear term is still unclear, we believe that it serves as a starting point for honeycomb lattice $t$-$J$ and has the potential to reveal the key mechanism for the emergence of SC order in honeycomb lattice $t$-$J$ and Hubbard
models.

In Sec.~\ref{sec:afm-order}, we study the AFM order
in the honeycomb lattice Hubbard model using mean field theory. At
half filling, the band structure has two Dirac cones, and the on-site
Coulomb repulsion favors a commensurate AFM order. Due
to the vanishing density of states of the Dirac cones, a finite
interaction strength is required to open an AFM gap on the Dirac
cones. At finite doping, the Dirac points grow into small pocket-like
Fermi surfaces. We first calculated the magnetic susceptibility and
show that the magnetic order is still commensurate. We then calculated
the phase diagram of the AFM phase in mean field approximation. A
highlight of the phase diagram is that at finite doping the AFM order
is suppressed at low temperature due to the fact that the commensurate
order does not gap the Fermi surface, and at large enough doping the
system reenters a paramagnetic state at low temperature while there is
an AFM phase at intermediate temperatures.
In Sec.~\ref{sec:coex-afm-superc} we study the coexistence of
AFM order and SC order using Ginzburg-Landau
theory. We first show that because of the symmetry of the honeycomb
lattice, a spin-singlet pairing and a spin-triplet pairing actually
has the same lattice symmetry transformation. Consequently the three
order parameters of AFM and spin-singlet and triplet SC can together
form a trilinear coupling term in the low energy effective
Hamiltonian. Therefore when there is a coexistence of AFM and SC
orders, the pairing naturally has both spin-singlet and spin-triplet
pairings. Moreover, in the presence of an AFM order, the trilinear
term becomes a quadratic coupling between two SC order parameters and
therefore the AFM order enhances SC order.
In Sec.~\ref{sec:tpt} we discuss the topological classification of
the three-order coexisting state. We first identify a possible
topological phase transition point where the quasiparticle gap
vanishes on one Dirac node. Then by calculating the change of Berry
phase connection near the nodal point across the phase transition, we
conclude that the Chern number of the SC state indeed changes across
the transition point and it separates two topologically different SC
states, which are topologically trivial and nontrivial respectively.

\section{Antiferromagnetic order}
\label{sec:afm-order}

In this section we study the AFM order in the honeycomb
lattice Hubbard model using mean field approximation. We start with
the following model,
\begin{equation}
  \label{eq:hub}
  H=-t\sum_{\langle ij\rangle \alpha}\left(c_{i\alpha}^\dagger c_{j\alpha}
    +\text{h. c.}\right)
  + U\sum_in_{i\uparrow}n_{i\downarrow}.
\end{equation}

\begin{figure}[htbp]
  \centering
  \includegraphics{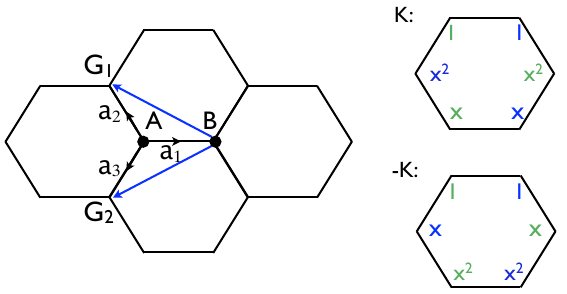}
  \caption{Structure of honeycomb lattice. In the left panel the three
    vectors $\bm a_i$ shows the direction of nearest neighbor bonds,
    and $\bm G_{1,2}$ are two primitive vectors of the triangular
    Bravais lattice. In the right panel the pairing symmetries at two
    Dirac points are shown, where $x=e^{i2\pi/3}$, and blue and green
    letters represent positive and negative phases respectively.}
  \label{fig:honeybs}
\end{figure}

The first term in the above Hamiltonian can be diagonalized in Fourier
space as the following,
\begin{equation}
  \label{eq:Etk}
  H_t=-t\sum_{k\alpha}\bar c_{k\alpha}^\dagger
  \begin{pmatrix}
    0 & \xi_k^\ast\\
    \xi_k & 0
  \end{pmatrix}\bar c_{k\alpha}, \end{equation}
where $\bar c_{k\alpha}=\begin{pmatrix}c_{Ak\alpha}&c_{Bk\alpha}\end{pmatrix}^T$
represents electron operators on two sites A and B in a unit cell (see Fig.~\ref{fig:honeybs}), and $\alpha=\uparrow,\downarrow$ denotes the electron spin. The function $\xi_k=1+e^{ik_1}+e^{ik_2}$, and $k_i=\bm k\cdot\bm a_i$ is the $i$-th
component of the momentum with respect to the two primitive
translation vector $\bm G_{1,2}$ of the triangular lattice, as shown
in Fig.~\ref{fig:honeybs}. (Here, the superscript $T$ denotes matrix transpose.) It is well-known that this represents a
band structure with two Dirac cones located at $\pm\bm
K=\pm\left(\frac{2\pi}3,-\frac{2\pi}3\right)$ (the momentum is given
in the reciprocal basis of $\bm G_{1,2}$). The second term provides an
on-site Coulomb repulsion and when $U$ is much greater than $t$ one
can restrict oneself in the single-occupied subspace and obtain a
$t$-$J$ model with antiferromagnetic interaction on nearest neighbor
bonds as a low-energy effective model. Hence at large enough $U$ the
system has an antiferromagnetic ground state.

Here we study this AFM order in mean field
approximation. We introduce the following SDW order parameter,
\begin{equation}
  \label{eq:Miz}
  \bm M_i=\left<\bm S_i\right>
\end{equation}
Plugging this into equation~(\ref{eq:hub}), the $U$ term can be decomposed into the following form in mean field approximation,
\begin{equation}
  \label{eq:u-mft}
  Un_{i\uparrow}\cdot n_{i\downarrow}
  =\frac U2n_i-2\bm M_i\cdot\bm S_i+\frac{\bm M_i^2}U
\end{equation}
Note, that the first term merely shifts the chemical potential of the system by $\frac U2$ and shall be ignored.

As discussed in Appendix~\ref{sec:comm-afm-order}, we consider a
commensurate order
\begin{equation}
  \label{eq:mz-comm}
  \bm M_i=(-)^iM_0\bm e_z
\end{equation}
where $(-)^i$ equals to 1 on sublattice A and $-1$ on sublattice B. With the mean field decomposition in equation~(\ref{eq:u-mft}), the Hamiltonian can be written in momentum space as
\begin{equation}
  \label{eq:H-mft-comm}
  H_{\text{MFT}}=\sum_{k\alpha}\bar{c}_{k\alpha}^\dagger
  \begin{pmatrix}
    -\mu+\alpha M_0 & -t\xi_k^\ast\\
    -t\xi_k & -\mu-\alpha M_0
  \end{pmatrix}
\bar{c}_{k\alpha}+\frac NUM_0^2,
\end{equation}

\begin{figure}[htbp]
  \centering
  \includegraphics{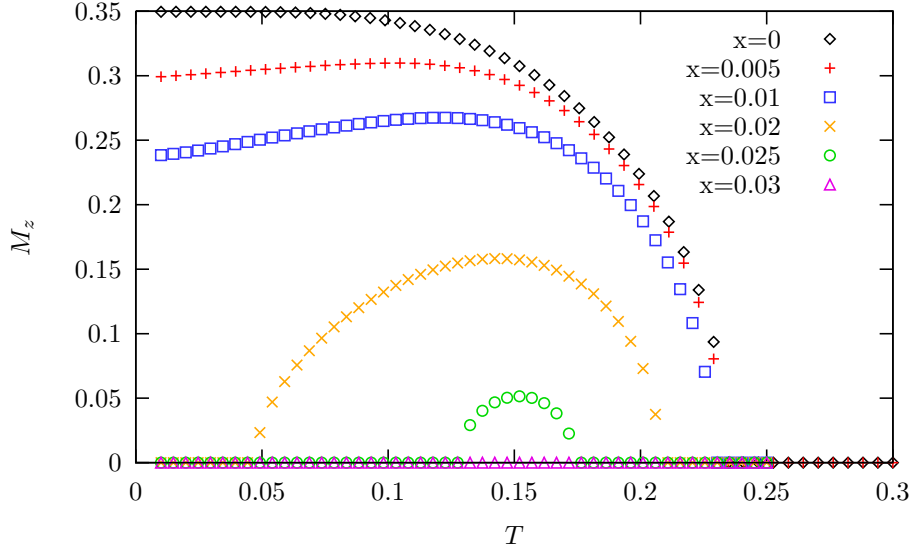}
  \caption{AFM order parameter as a function of
    temperature at different doping levels. The plot was calculated
    with $U=3t$. Both $T$ and $M_z$ axes are in units of $t$.}
  \label{fig:Tscan_overlay}
\end{figure}

\begin{figure}[htbp]
  \centering
  \includegraphics{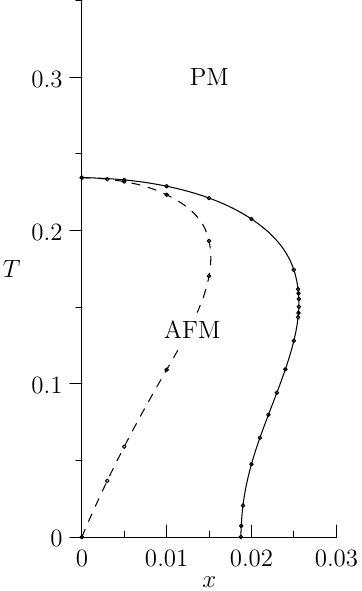}
  \caption{Mean field phase diagram of the Hubbard model. The solid line
    is where the AFM order parameter $M_z$ vanishes, and
    it separates the AFM phase and the
    paramagnetic (PM) phase. The dashed line is where $M_z=\mu$, and
    it separates the two superconducting phases with different
    topological classifications in the weak coupling limit. This is
    discussed in Sec.~\ref{sec:tpt}.}
  \label{fig:afpd}
\end{figure}

Using the Hamiltonian described in equation~\eqref{eq:H-mft-comm}, we plot
the mean field phase diagram through numerically minimizing the
Hamiltonian with respect to the AFM order parameter
$M_z$ at a fixed doping $x$. Results of $M_z$ as a function of
temperature at different doping levels are plotted in
Fig.~\ref{fig:Tscan_overlay}, and the phase diagram determined from
this self-consistent calculation is plotted in Fig.~\ref{fig:afpd}.

At zero doping, the $M_z$ curve has a typical parabolic shape, showing
a paramagnetic high temperature phase and an antiferromagnetic low
temperature phase separated by a continuous phase transition.
At finite doping, the AFM order is generally suppressed
as the commensurate order cannot gap the Fermi surface. The
suppression is stronger at low temperature and weaker at high
temperature, since at high temperature the Fermi surface is not quite
clear when $T\geq T_F$. At doping levels $x=0.02$ and $x=0.025$, the
magnetic order is completely suppressed at low temperatures and the
system reenters the paramagnetic phase at a lower critical
temperature. For these two dopings, the antiferromagnetic phase only
exists between two critical temperatures. At the doping level
$x=0.0256$, the antiferromagnetic phase disappears as the two critical
temperatures merge. Of course, according to the Mermin-Wagner theorem, the AFM order will be killed by quantum fluctuation at finite temperature for strictly 2D systems. However, for realistic material, the interlayer coupling will always stabilize AFM order at finite temperature. Therefore, the above phase diagram is still reasonable for realistic systems and can be improved by considering both quantum fluctuations and interlayer couplings.

\section{Coexistence of three orders}
\label{sec:coex-afm-superc}

In the previous section, we see that the Hubbard model on the honeycomb
lattice develops commensurate AFM order at zero and
small dopings. In this section, we argue that this AFM
order will induce superconducting order with mixed singlet and triplet
pairings.

\begin{figure}[htbp]
  \centering
  \subfigure[\label{fig:pairing:singlet} Spin-singlet pairing.]
  {\includegraphics{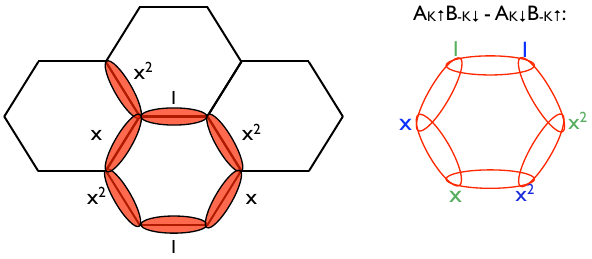}}
  \subfigure[\label{fig:pairing:triplet} Spin-triplet pairing.]
  {\includegraphics{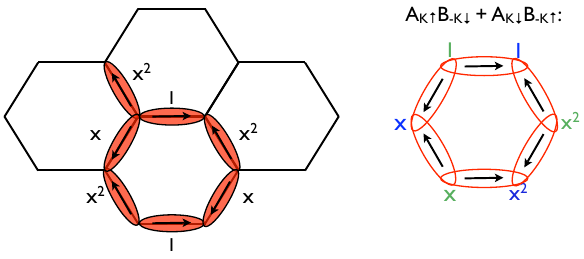}}
  \caption{Spin-singlet and spin-triplet pairing symmetry. The complex
    pairing amplitudes are noted along the bonds, where $x=e^{i2\pi/3}$.}
  \label{fig:pairing}
\end{figure}

One interesting feature observed in the numerical study of
Ref.~\onlinecite{Gu2011} is that the superconducting state has both
spin-singlet and spin-triplet pairings. In a lattice with inversion
symmetry, singlet and triplet pairing order parameters have even and
odd parity under inversion symmetry operation respectively and
therefore do not mix. However, the honeycomb lattice does not have
inversion symmetry and therefore in general allows the mixing of
singlet and triplet pairing order parameters. Both the singlet and
triplet pairing order parameters found in the aforementioned numerical
study have a 120-degree spatial pattern: The phase on the bonds of
the honeycomb lattice rotates by 120 degrees around the center of each
hexagon, as shown in Fig.~\ref{fig:pairing}. The singlet pairing
symmetry is the same as the $d+id$ pairing obtained in other
researches~\cite{Honerkamp2008,Wu2013}.

The same spatial pattern of the two pairing symmetries implies the
mixing of spin-singlet and spin-triplet pairing in the presence of
AFM order. Since the spin-singlet and spin-triplet
pairings have the same spatial pattern, they transform in the same way
under three-fold rotation. Therefore it is easy to check that the
following combination of the three order parameters is invariant under
all symmetry transformations including spin rotation, time reversal
lattice symmetry transformations, and electromagnetic U(1) gauge
symmetry transformation,
\begin{equation}
  \label{eq:H-tri-linear}
  H_3=\lambda_3 \bm M\cdot\Delta_s^\ast\bm \Delta_t+\text{h. c.},
\end{equation}
and therefore is allowed to appear in the low-energy effective
Hamiltonian of the system. In Eq.~\eqref{eq:H-tri-linear} $\Delta_s$
and $\bm \Delta_t$ denote the superconducting order parameter of
spin-singlet and spin-triplet pairing respectively, where the latter
is a spin-1 vector. The presence of this trilinear term implies that
once two of the three order parameters become nonzero, the third one
will be automatically induced, as the symmetry that the third order
breaks has already been broken by the other two orders. Therefore in the
honeycomb lattice if there is a coexisting state of AFM and SC, the SC
order parameter naturally contains both spin-singlet and spin-triplet
components.

Moreover, the trilinear term also implies that the presence of AFM
order helps the formation of SC order. In an AFM state, one
can replace the $\bm M$ order parameter by its expectation value and
the trilinear term in Eq.~\eqref{eq:H-tri-linear} becomes a quadratic
term that couples the two SC order parameters $\Delta_s$ and
$\bm\Delta_t$. The sign of the trilinear term will determine the
relative orientation of the AFM order parameter and the d-vector of
the triplet pairing, but the resulting quadratic term always favors SC
ordering. In the rest of this section we study this effect using a
concrete model.

At mean field level, the onsite repulsive interaction in the Hubbard
model cannot be decomposed in the superconducting channel. Therefore a
naive mean field analysis of the Hubbard model does not reveal a
superconducting order. However, we expect that in the Mott insulating
phase the onsite repulsive interaction introduces a nearest-neighbor
Heisenberg interaction through second-order virtual processes, and
this interaction can lead to SC order. Hence in this section
we only calculate the susceptibility of the superconducting operator
from the kinetic energy. Once the susceptibility diverges as
$T$ goes to zero, a superconducting order will raise once we add the
appropriate interaction.

Our goal is to study the quadratic terms of the superconducting order
parameter in the Hamiltonian,
\begin{equation}
  \label{eq:tri-linear}
  H_{\text{quad}}=\frac12\lambda_{ab}\Delta_a\Delta_b,
\end{equation}
where $a,b=s,t$ stands for singlet and triplet pairings, respectively.
Here, we only consider the $z$ component of the triplet pairing, and use
$\Delta^t$ to denote $\Delta_t^z$, since we assume the magnetization is in the $z$ direction, which only couples to $\Delta_t^z$ through the trilinear term in Eq.~\eqref{eq:H-tri-linear}.
To study the superconducting order induced by antiferromagnetism, we
assume that there is an AFM order parameter calculated
self-consistently from the mean field Hamiltonian, and study the coupling
constant $\lambda$ in equation~\eqref{eq:tri-linear} diagrammatically.
We use only the kinetic energy term in equation~\eqref{eq:u-mft},
and add the coupling between the SC order parameters and the electrons,
\begin{equation}
  \label{eq:H-afm}
  H=\sum_{k\alpha}\bar c_{k\alpha}^\dagger T_{k\alpha}\bar c_{k\alpha}
  +\frac23\Delta_s^\ast \sum_k
  \bar c_{k\uparrow}^T\Gamma^s_k\bar c_{-k\downarrow}
  +\frac23\Delta_t^\ast\sum_k
  \bar c_{k\uparrow}^T\Gamma^t_k\bar c_{-k\downarrow}
  +\text{h. c.},
\end{equation}
where $\bar c_{k\alpha}=(c_{Ak\alpha},c_{Bk\alpha})^T$, and the matrices $T_k$ and
$\Gamma_{s,t}$ are defined as the following,
\begin{equation}
  \label{eq:tkg}
  T_{k\alpha}=\begin{pmatrix}
    -\mu +\alpha M^z&-t\xi_k\\
    -t\xi_k^\ast&-\mu-\alpha M^z
  \end{pmatrix},\quad
  \Gamma^{s,t}_k=\begin{pmatrix}
    0&\gamma_k\\\pm\gamma_{-k}&0
  \end{pmatrix},
\end{equation}
where $\gamma_k=1+e^{-i(k_1+2\pi/3)}+e^{-i(k_2+4\pi/3)}$.
We notice that $\Gamma_s(-\bm k)=\Gamma_s(\bm k)^T$, and $\Gamma_t(-\bm k)=-\Gamma_t(\bm k)^T$. Thus, $\Delta_{s,t}$ couples to electron pairings $c_{k\uparrow}c_{-k\downarrow}\mp c_{-k\uparrow}c_{k\downarrow}$, respectively, consistent with the singlet and triplet pairing symmetries. As we discussed before, here we only consider the $z$ component of the vector $\bm\Delta_t$, which couples to electron operators in the following general form,
$\bm\Delta_t\cdot c_{k\alpha}i\sigma^y_{\beta\gamma}\bm\sigma_{\gamma\delta}c_{-k\delta}$.
Therefore, the $z$ component of $\bm \Delta_t$ couples to the symmetric pairing channel $c_{k\uparrow}c_{-k\downarrow}+ c_{-k\uparrow}c_{k\downarrow}$.

From this effective Hamiltonian, the coefficient $\lambda$ can be
calculated as following,
\begin{equation}
  \label{eq:lambda-diag}
  \lambda_{ab}=-\frac1{\beta V}\sum_{\omega_n}\sum_k\tr\left[
    \frac23\Gamma^a_k
    G_\uparrow(k, i\omega_n)
    \frac23(\Gamma^b_k)^\dagger
    G_\downarrow(-k, -i\omega_n)
    \right],
\end{equation}
where the Green's function $G_\alpha(k,i\omega_n)$ is derived from the
first term in equation~\eqref{eq:H-afm},
\begin{equation}
  \label{eq:gkw}
  G_\alpha(k, i\omega_n)
  =(i\omega_n-T_{k\alpha})^{-1}.
\end{equation}

Plugging equation~\eqref{eq:gkw} into equation~\eqref{eq:lambda-diag}, we
get the following result after some manipulations,
\begin{gather}
  \label{eq:lambda-sumkw}
  \lambda_{st}=-\frac{16}9\mu M^z
  \frac1{\beta V}\sum_{k,\omega_n}
  \frac{|\gamma_k|^2}{[(i\omega_n+\mu)^2-E_k^2]
    [(-i\omega_n+\mu)^2-E_k^2]},\\
  \label{eq:lambda-ss}
  \lambda_{ss,tt}=\frac49\frac1{\beta V}\sum_{k,\omega_n}
  \frac{(|\gamma_k|^2+|\gamma_{-k}|^2)(\omega_n^2+\mu^2+(M^z)^2)
    \pm2\gamma_k^\ast\gamma_{-k}t^2|\xi_k|^2}
  {[(i\omega_n+\mu)^2-E_k^2][(i\omega_n-\mu)^2-E_k^2]}.
\end{gather}
where $E_k=\sqrt{(M^z)^2+|\xi_k|^2}$ is the quasiparticle energy. Now
we can evaluate the frequency summation and get
\begin{equation}
  \label{eq:lambda-sumk}
  \begin{split}
 \lambda_{st}=\frac{16}9\mu M^z
  \int\frac{d^2k}{(2\pi)^2}|\gamma_k|^2
  \left[
    \frac1{8\mu E_k(E_k+\mu)}
    (2n_F(E_k+\mu)-1)
    -\right.\\
  \left.
    \frac1{8\mu E_k(E_k-\mu)}
    (2n_F(E_k-\mu)-1)
  \right]
  \end{split},
\end{equation}
where $n_F(z)=(e^{\beta z}+1)^{-1}$ is the Fermi occupation number,
and
\begin{equation}
  \label{eq:lambda-sumk-sstt}
  \begin{split}
    \lambda_{ss,tt}=\frac{4}9
    \int\frac{d^2k}{(2\pi)^2}
    \left[
      \frac{2\mu E_k(|\gamma_k|^2+|\gamma_{-k}|^2)
      +|\xi_k|^2|\gamma_k\pm\gamma_{-k}|^2}{8\mu E_k(E_k+\mu)}
      (2n_F(E_k+\mu)-1)
      -\right.\\
    \left.
      \frac{-2\mu E_k(|\gamma_k|^2+|\gamma_{-k}|^2)
      +|\xi_k|^2|\gamma_k\pm\gamma_{-k}|^2}{8\mu E_k(E_k-\mu)}
      (2n_F(E_k-\mu)-1)
    \right]
  \end{split},
\end{equation}

\begin{figure}[htbp]
  \centering
  \includegraphics{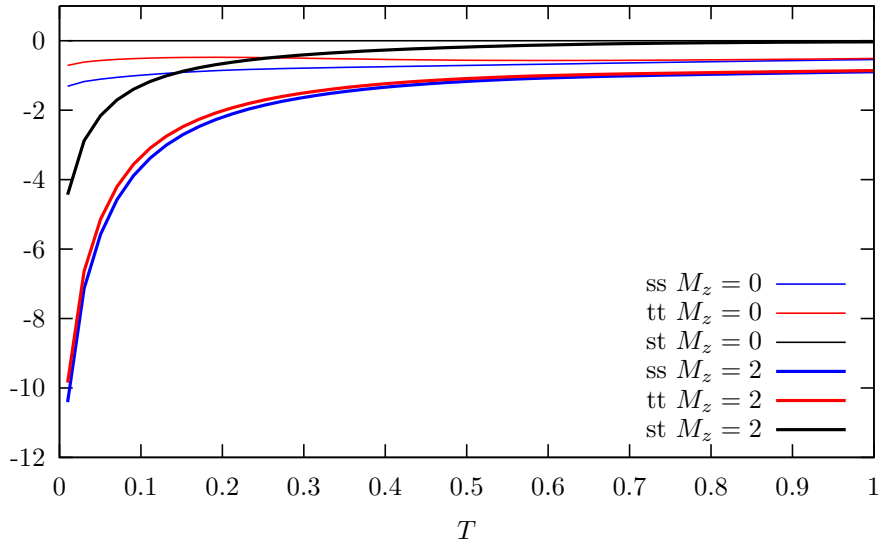}
  \caption{Plot of $\lambda_{ss}$, $\lambda_{tt}$, and
    $\lambda_{st}$. The system is at doping $x=0.05$.}
  \label{fig:wwomag1}
\end{figure}

Now we show some plots of $\lambda$ calculated from
equations~\eqref{eq:lambda-sumk} and~\eqref{eq:lambda-sumk-sstt}. In
Fig.~\ref{fig:wwomag1} we show $\lambda_{ss}$, $\lambda_{tt}$ and
$\lambda_{st}$ at doping $x=0.05$ with and without a magnetic gap. In
the plot we see that without magnetic gap, $\lambda_{ss}$ and
$\lambda_{tt}$ (black diamonds and red crosses) are flat at high
temperatures and only diverge at $T\ll T_F$. Also without a magnetic
order $\lambda_{st}=0$ (this is not shown in the plot, but we know
this because a nonvanishing $\lambda_{st}$ in the absence of magnetic
order would break spin rotation symmetry). Hence without magnetic
order, the system is going superconducting only when it is cooled down
below Fermi temperature. With magnetic gap, however, $\lambda_{ss}$, $\lambda_{tt}$, and $\lambda_{st}$ (blue squares,
yellow crosses, and green circles) all diverge in a similar manner at much
higher temperature, showing a tendency towards SC order at
temperature even higher than the Fermi temperature.
We notice that, in addition to the susceptibility, the interaction strength also affects the SC transition temperature.
Here, we assume that the interaction strength, arising from virtual antiferromagnetic spin exchanges in the limit of $U\gg t$, does not have a strong dependence on doping.
Therefore, the interaction strength can be treated as a constant across the antiferromagnetic transition point.
Comparing to the
case without magnetic order, we conclude that this SC order
is induced by the AFM order.

\begin{figure}[htbp]
  \centering
  \includegraphics{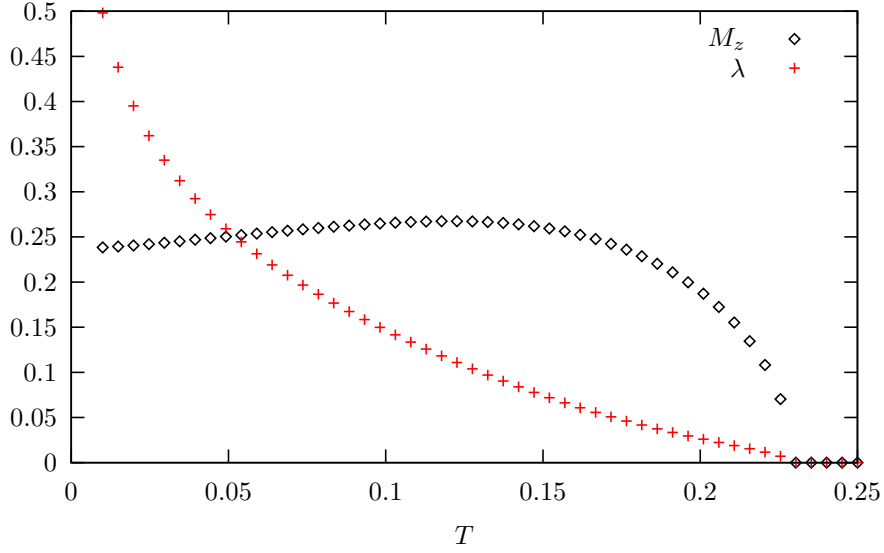}
  \caption{Plot of $\lambda_{st}$ as a function of temperature at doping $x=0.02$.}
  \label{fig:trilin-run3}
\end{figure}

\begin{figure}[htbp]
  \centering
  \includegraphics{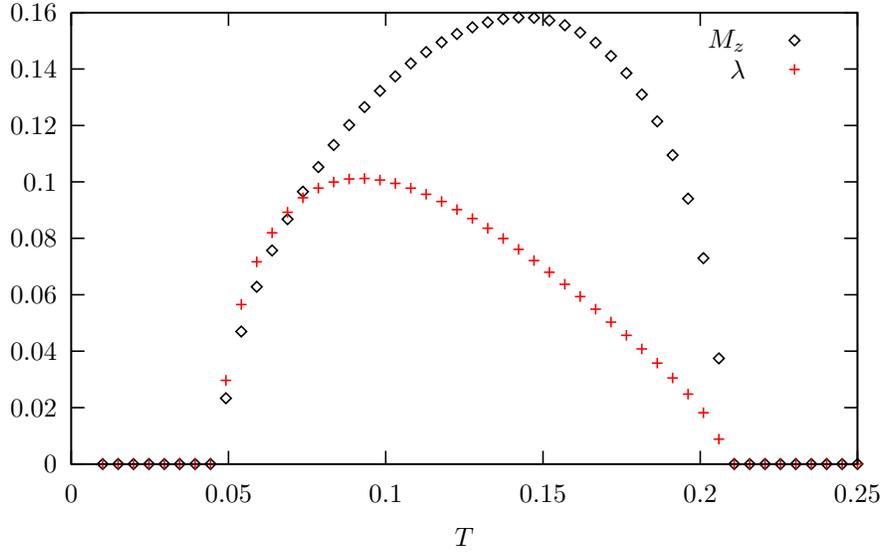}
  \caption{Plot of $\lambda_{st}$ as a function of temperature at doping $x=0.025$.}
  \label{fig:trilin-run4}
\end{figure}

Then we show some plots of $\lambda_{st}$ calculated from
equation~\eqref{eq:lambda-sumk} with magnetic order calculated
self-consistently. In Figs.~\ref{fig:trilin-run3} and
\ref{fig:trilin-run4} we plot $\lambda_{st}$ as a function of
temperature at certain doping levels. The calculation is based on the
mean field result of $M_z$ shown in Fig.~\ref{fig:Tscan_overlay}. At
$x=0.02$, in the antiferromagnetic phase $\lambda$ increases as
temperature drops and eventually diverges as $T$ goes to zero. At
$x=0.025$, $\lambda$ also increases as temperature drops when first
entering the antiferromagnetic phase, but $\lambda$ eventually drops
to zero as the magnetic order disappears at lower temperature.

In summary, in this section we see that on the honeycomb lattice, a trilinear
term that couples the AFM order and two SC orders of different pairing
symmetries is allowed by symmetry and in general exists in the
effective Hamiltonian. This term induces SC order in the AFM
phase. This argument qualitatively explains the three-order coexisting
phase observed in the numerical study~\cite{Gu2011}.

\section{Topological phase transition}
\label{sec:tpt}

In this section we study the topological classification of the
coexisting order phase discussed in
Sec.~\ref{sec:coex-afm-superc}. This phase has both superconducting
and AFM orders, and therefore it has neither time
reversal nor U(1) charge symmetry and such systems in two dimensions
are classified by an integer topological invariant~\cite{Kitaev2009},
which can be calculated from the Chern number of the Bogolyubov-de
Gennes (BdG) Hamiltonian~\cite{Qi2010b}.

One interesting feature of the coexisting order state is that it can
be either topologically trivial or nontrivial in different parameter
ranges, and there is a topological phase transition separating the two
regimes. We start with identifying this topological phase transition
in the phase diagram. Analogous to topological insulators, topological
superconductors have gapped fermionic quasiparticle excitations
described by a gapped BdG Hamiltonian, and it cannot be smoothly tuned
to a topologically trivial state without closing the gap of
quasiparticle excitations, or the superconducting gap. Hence a
necessary condition of a topological phase transition is the closing
of the quasiparticle gap.

Without losing generality, in this section we assume the SC pairing is
in the weak coupling limit, or the SC gap is much smaller than the AFM
gap. In this limit, we first study the AFM state using mean field
theory as in Sec.~\ref{sec:afm-order} and obtain the band
structure with a AFM mean field gap $M_z$. Secondly, as discussed in
Sec.~\ref{sec:coex-afm-superc}, the AFM order induces a SC order
with coexisting spin-singlet and spin-triplet pairings. Here to
discuss the topological classifications and the topological phase
transition, we only consider a weak SC pairing on top of the mean
field band structure of the AFM state and ignore the feedback of the
SC order on the AFM order parameter. For superconductors in the weak
coupling limit, their topological classification is determined by the
normal state band structure and pairing symmetry. In our case, the
topological classification of superconducting states with coexisting
spin-singlet and spin-triplet pairing symmetries is determined by the
mean field band structure of the AFM state.

In the coexisting order phase, the quasiparticle gap indeed closes at
a particular point in the phase diagram, because the spin-singlet and
spin-triplet superconducting order parameters have nodes at one of the
two Dirac cones. From the form of the gap function in
Eq.~\eqref{eq:tkg} we can see that the gap functions take the
following form at the two Dirac points $k=\pm K$,
\begin{equation}
  \label{eq:Delta_st_pmK}
  \Gamma_K^{s,t}=
  \begin{pmatrix}
    0 & 0 \\ \pm1 & 0
  \end{pmatrix},\quad
  \Gamma_{-K}^{s,t}=
  \begin{pmatrix}
    0&1\\0&0
  \end{pmatrix}.
\end{equation}
This means that in both pairing symmetries, the A sublattice state at
$K$ is paired up with the B sublattice state at $-K$, while the B
sublattice state at $K$ is \emph{not} paired up with the A sublattice
state at $-K$. It can be simply understood from the Bloch
wavefunctions: As shown in Fig.~\ref{fig:pairing}, the $A_K$-$B_{-K}$
pairing immediately leads to the 120 degree pattern, whereas the
$B_K$-$A_{-k}$ pairing leads to the $-120$ degree pattern. Hence the
superconducting gap function vanishes at the latter point if we take
the 120 degree pairing pattern. When the SC order coexists with the
AFM order, the total quasiparticle gap is the sum of the SC gap and
AFM gap. Consequently the quasiparticle gap vanishes if the AFM gap
vanishes at the Dirac nodes, which happens when the Fermi level
touches the bottom of the band in the AFM state, or $\mu=M_z$ as shown
in Eq.~\eqref{eq:H-mft-comm}.

Next, we argue that the superconducting state indeed goes through a
topological phase transition when the gap opens a node at $\mu =
M_z$. At the transition point, the gap function vanishes for pairing
between the A sublattice state at $K$ and B sublattice state at $-K$,
while other states remain gapped. Hence across the transition point
the change in the Chern number comes from the change of the Berry
curvature of the A sublattice states near $K$ and B sublattice states
at $-K$. To calculate this change we can use a simplified model of
these states. Considering only the spin-up states of the A sublattice
near $K$ and spin-down states of B sublattice near $-K$, we can expand
the mean field Hamiltonian in Eq.~\eqref{eq:H-afm} and get the
following effective two-band BdG Hamiltonian,
\begin{equation}
  \label{eq:HBdG2}
  H_{\text{eff}}=
  \begin{pmatrix}
    c_{\delta kA\uparrow}^\dagger & c_{\delta kB\downarrow}
  \end{pmatrix}
  \begin{pmatrix}
    -\mu+M_z+t\delta k^2 & (\Delta^s+\Delta^t)(\delta k_x+i\delta k_y)\\
    (\Delta^s+\Delta^t)(\delta k_x-i\delta k_y) & \mu-M_z-t\delta k^2
  \end{pmatrix}
  \begin{pmatrix}
    c_{\delta kA\uparrow} \\ c_{\delta kB\downarrow}^\dagger
  \end{pmatrix},
\end{equation}
where $\delta\bm k = \bm k-\bm K$ is the momentum measured from the
Dirac point $K$, and $\delta k_x$ and $\delta k_y$ are two orthogonal
components of $\delta\bm k$. The Chern number of this simplified BdG
Hamiltonian is calculated in Ref.~\onlinecite{Qi2010b}, and it is
topologically trivial if $\mu<M_z$, and it has a nontrivial Chern
number of two if $\mu>M_z$. From this simplified model we conclude
that at the transition point of $\mu=M_z$, the total Chern number of
the system changes by two, and therefore it is indeed a topological
phase transition separating two different superconducting states with
different topological classifications. The change in Chern number can
be obtained from an effective model near the nodal point, but the
total Chern number of the complete BdG Hamiltonian can only be
determined by integrating over the full Brillouin zone and summing
over all bands. However, using a simple argument we can see that the
state of $\mu<M_z$ is indeed topologically trivial with Chern number
zero, because one can smoothly connect this state to vacuum state but
sending $M_z$ to infinite without closing the quasiparticle
gap. Therefore the superconducting state at the other side of the
transition, with $\mu>M_z$, must be a topologically nontrivial state
with Chern number equal to two. This result can be checked by
calculating the Chern number using the full mean field Hamiltonian in
Eq.~\eqref{eq:H-afm}.

In the weak coupling limit, the sign of $\mu-M_z$ can be calculated
self-consistently using the mean field theory described in
Sec.~\ref{sec:afm-order} as we ignore the feedback of
SC order on the AFM order. The phase boundary of the aforementioned
topological phase transition is plotted in Fig.~\ref{fig:afpd} by a
dashed line. The region enclosed by the dashed line has $\mu<M_z$ and
the SC state is topologically trivial, and the region between the
dashed line and the solid line has $\mu>M_z$ and the SC state is
topologically nontrivial.

\section{Conclusions}
\label{sec:conclusions}

In this work we study the AFM and SC orders in the doped Hubbard model
on the honeycomb lattice. A phase diagram of the AFM order is obtained by
self-consistent mean field calculation, and a commensurate AFM order
is found at low temperature and small dopings. Using symmetry
analysis, we show that a trilinear term that couples together AFM
order and both spin-singlet/spin-triplet SC orders is allowed by
symmetry, and such a term implies that the AFM order induces the two
SC orders and gives rise to a phase with coexisting AFM and SC orders
with both pairing symmetries. At last, we show that the three-order
coexisting phase is separated by a topological phase transition to a
topologically trivial SC phase and a topologically nontrivial SC
phase with Chern number equals to two.

Of course, it will be of great interest to examine the proposed effective field theory
in experiment. The recently discovered spin $1/2$ honeycomb lattice
Mott-insulator InV$_{1/3}$Cu$_{2/3}$O$_3$~\cite{InVCuO} would be an
appealing candidate if it could be doped experimentally. The recent ultra cold Fermi gas in the honeycomb optical lattice~\cite{Tarruell2012}
is another way to realize the honeycomb lattice $t-J$ model.

\begin{acknowledgments}
We would like to thank Dun-Hai Lee, Fa Wang, and Hong Yao for helpful discussions. The work at MIT was supported by DOE Office of Basic Energy Sciences, Division of Materials Sciences and Engineering under Award de-sc0010526. Z.C.G. acknowledges Direct Grants No. 4053224 and No. 4053409 from The Chinese University of Hong Kong and funding from Hong Kong Research Grants Council (ECS No. 2191110 and No. 24301516).
\end{acknowledgments}

\appendix

\section{Commensurability of the AFM order}
\label{sec:comm-afm-order}

Here we study the commensurability of the antiferromagnetic order in
the system. In the large-$U$ limit, superexchange processes create an
antiferromagnetic interaction between nearest-neighbor spins. At
half filling, the ground state of the Heisenberg model with
nearest-neighbor interaction is a commensurate N\'eel order with
antiparallel spins on two sublattices. After doping, the
AFM order may become incommensurate, as it does on a
square lattice. In this section we study this possibility through
evaluating the spin susceptibility. The peak momentum of the
susceptibility will point out the commensurability of the order.

We consider the following static spin susceptibility at a finite
wave vector $\bm Q$, which is defined as
\begin{equation}
  \label{eq:chi-def}
  \chi^{ij}_{ab}(\bm Q, \omega=0)
  =\int d\tau \sum_{kk^\prime}\bar c_{k+Q}^\dagger(\tau)
  \sigma^i\otimes\mu_a \bar c_k(\tau)
  \bar c_{k^\prime-Q}^\dagger(0)
  \sigma^j\otimes\mu_b \bar c_{k^\prime}(0),
\end{equation}
where $a,b=A$ or $B$ denotes the two sublattices, and the matrices
$\mu_a$ are defined as
\begin{equation}
  \label{eq:mu-AB}
  \mu_A=
  \begin{pmatrix}
    1&0\\0&0
  \end{pmatrix},\quad
  \mu_B=
  \begin{pmatrix}
    0&0\\0&1
  \end{pmatrix}.
\end{equation}

Without losing generality, we consider $\chi^{zz}$, which can be
evaluated using Green's function as following
\begin{equation}
  \label{eq:chi-eval}
  \chi_{ab}^{zz}(\bm Q)
  =\frac1\beta\sum_{\omega_n} \sum_{k}
  \tr\left(\bar G(\bm k+\bm Q, \omega_n)\sigma^z\otimes\mu_a
  \bar G(\bm k, \omega_n) \sigma^z\otimes\mu_b\right),
\end{equation}
where $\bm G$ is derived from the kinetic energy in
equation~\eqref{eq:hub},
\begin{equation}
  \label{eq:Gbar}
  \bar G^{-1}(\bm k, \omega_n)
  =i\omega_n-
  \begin{pmatrix}
    -\mu&-t\xi_k^\ast\\
    -t\xi_k&-\mu
  \end{pmatrix}.
\end{equation}

First, we study the spin susceptibility between two sublattices
$\chi_{AB}^{zz}$. At zero temperature and assuming $\mu\geq0$,
equation~\eqref{eq:chi-eval} becomes
\begin{equation}
  \label{eq:xi-intk}
  \begin{split}
    \chi_{AB}^{zz}(\bm Q)=
    &-\sum_k\frac{\xi_{k+Q}^\ast
      \xi_k}{2|\xi_k||\xi_{k+Q}|(|\xi_k|+|\xi_{k+Q}|)}\\
    &-\sum_k\frac{\xi_{k+Q}^\ast
      \xi_k[|\xi_{k+Q}|\theta(\mu-|\xi_k|)
      -|\xi_k|\theta(\mu-|\xi_{k+Q}|)]}
    {2|\xi_k||\xi_{k+Q}|(|\xi_k|+|\xi_{k+Q}|) (|\xi_k|-|\xi_{k+Q}|)}.
  \end{split}
\end{equation}

Similarly for the susceptibility of the same sublattice, we get
\begin{equation}
  \label{eq:xi-intk2}
  \begin{split}
    \chi_{AA}^{zz}(\bm Q)=
    &\sum_k\frac1{2(|\xi_k|+|\xi_{k+Q}|)}\\
    &+\sum_k\frac{|\xi_k|\theta(\mu-|\xi_k|)
      -|\xi_{k+Q}|\theta(\mu-|\xi_{k+Q}|)}
    {2(|\xi_k|+|\xi_{k+Q}|) (|\xi_k|-|\xi_{k+Q}|)}.
  \end{split}
\end{equation}

\begin{figure}[htbp]
  \centering
 \subfigure[Intersublattice susceptibility $|\chi_{AB}^{zz}|$.]
 {\includegraphics{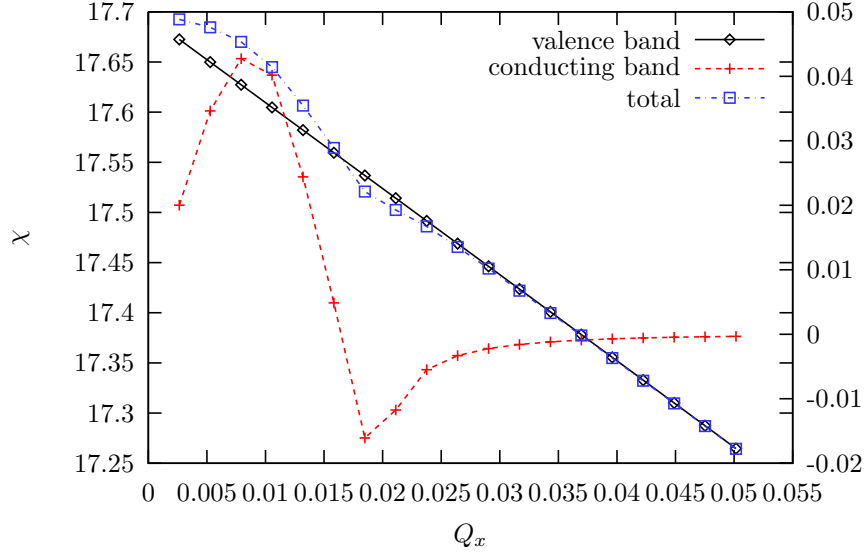}}
 \subfigure[Intrasublattice susceptibility $\chi_{AB}^{zz}$.]
 {\includegraphics{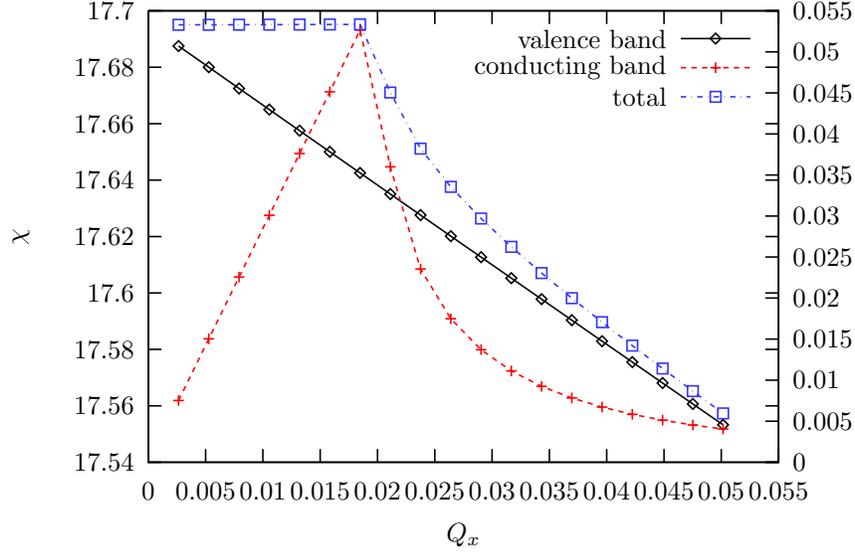}}
  \caption{Inter- and intrasublattice susceptibility. The
    conducting-band contribution shown by red curves is scaled
    differently from the other curves: The former uses the scale on
    the right and the latter uses the scale on the left.}
  \label{fig:suscep}
\end{figure}

\begin{figure}[htbp]
  \centering
  \includegraphics{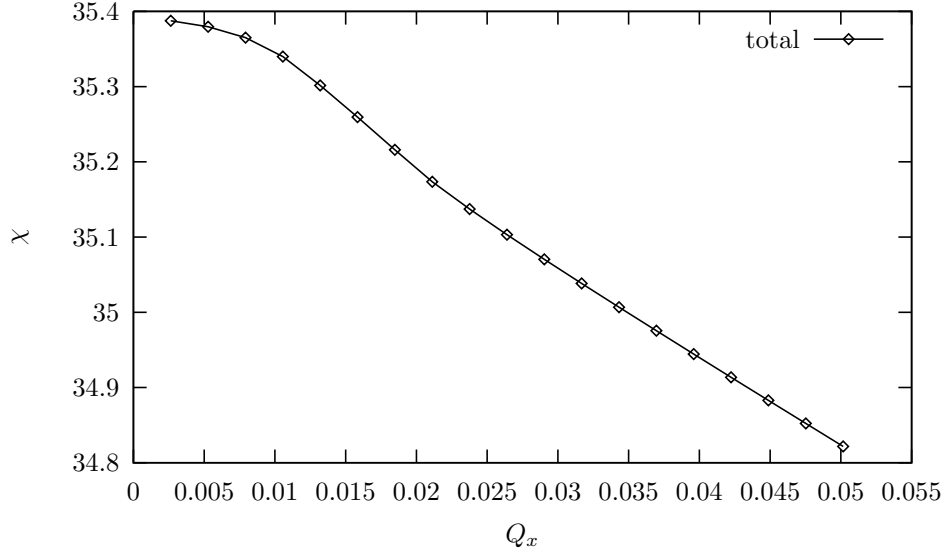}
  \caption{Total susceptibility $|\chi_{AB}^{zz}|+\chi_{AB}^{zz}$.}
  \label{fig:suscep-total}
\end{figure}

\begin{figure}[htbp]
  \centering
  \includegraphics{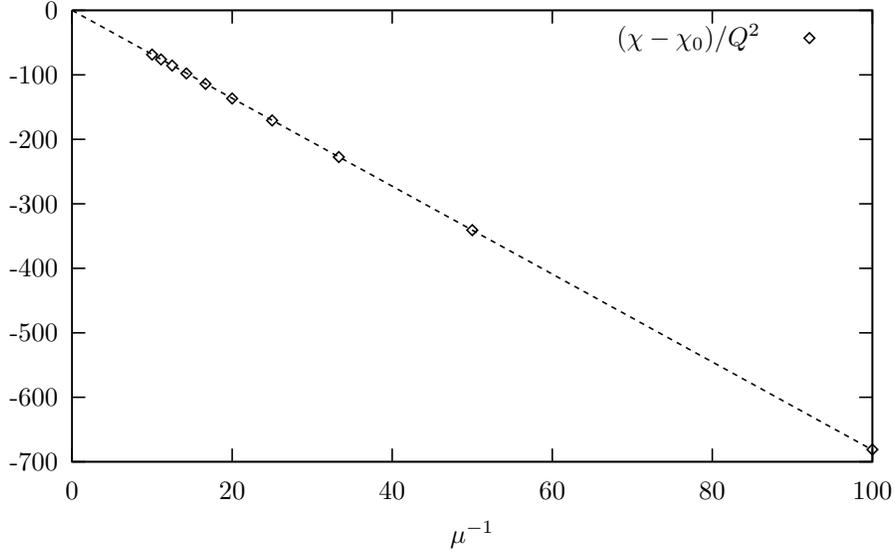}
  \caption{Quadratic term in $|\chi_{AB}^{zz}|$.}
  \label{fig:quad}
\end{figure}

The susceptibility obtained in equation~\eqref{eq:xi-intk} and
\eqref{eq:xi-intk2} can be separated into two terms: The first two
terms in the two equations come from the filled valence band, and the
second two terms come from the conducting band in which the Fermi
level sits. The contribution from the valence band does not depend on
doping and has a maximum at commensurate wave vector, while the
contribution from the conducting band has a maximum at incommensurate
wave vector which connects the two sides of the Fermi surface. The
intersublattice and intrasublattice susceptibilities are ploted as a
function of $\bm Q$ in Fig.~\ref{fig:suscep}. As discussed before,
the contribution from valence band and conducting band has maxima at
commensurate and incommensurate wave vectors respectively, but the
total susceptibility peaks at $(0,0)$ for the intersublattice case,
and the intrasublattice susceptibility is almost level near $(0,0)$
but it is slightly higher at incommensurate position. When added
together, the total susceptibility favors commensurate susceptibility,
as shown in Fig.~\ref{fig:suscep-total}.
In fact, this behavior is observed at different values of $\mu$, as Fig.~\ref{fig:quad} shows that $\chi_{AB}^{zz}$ has a maximum at $Q=0$ for all values of $\mu$.
Because of this result, we
only consider commensurate AFM order in the main text.

\bibliography{Honeycomb}

\end{document}